\documentclass[preprint,superscriptaddress,amsmath,amssymb,prl,floatfix,endfloats*]{revtex4}

\usepackage{graphicx}
\usepackage{amssymb}
\usepackage{dcolumn}
\usepackage{bm}

\begin{document}
\title{\textit{In situ} observation of strongly interacting ferromagnetic domains in a shaken optical lattice}
\author{Colin V. Parker}
\affiliation{The James Franck Institute and Department of Physics, University of Chicago, Chicago, IL 60637, USA}
\author{Li-Chung Ha}
\affiliation{The James Franck Institute and Department of Physics, University of Chicago, Chicago, IL 60637, USA}
\author{Cheng Chin}
\affiliation{The James Franck Institute and Department of Physics, University of Chicago, Chicago, IL 60637, USA}
\affiliation{Enrico Fermi Institute, University of Chicago, Chicago, IL 60637, USA}
\date{\today}



\maketitle
{\bf Solid state systems derive their richness from the interplay between interparticle interactions and novel band structures that deviate from those of free particles.
Strongly interacting systems, where both of these phenomena are of equal importance, exhibit a variety of theoretically interesting and practically useful phases.
Systems of ultracold atoms are rapidly emerging as precise and controllable simulators, and it is precisely in this strongly interacting regime where simulation is the most useful.
Here we demonstrate how to hybridize Bloch bands in optical lattices to introduce long-range ferromagnetic order in an itinerant atomic system.
We find spontaneously broken symmetry for bosons with a double-well dispersion condensing into one of two distinct minima, which we identify with spin-up and spin-down.
The density dynamics following a rapid quench to the ferromagnetic state confirm quantum interference between the two states as the mechanism for symmetry breaking.
Unlike spinor condensates, where interaction is driven by small spin-dependent differences in scattering length\cite{Stenger:1998p1770,Ho:1998p1766,Sadler:2006p1742}, our interactions scale with the scattering length itself, leading to domains which equilibrate rapidly and develop sharp boundaries characteristic of a strongly interacting ferromagnet.}

In pursuit of strongly interacting ultracold atomic systems, much effort has focused on exploiting or engineering interactions capable of generating long-range ordered phases.
Efforts to simulate ferromagnetism using spinor Bose gases\cite{Stenger:1998p1770,Chang:2005p1771,StamperKurn:2012p1756} have taken advantage of contact and weak dipole interactions to form domains\cite{Sadler:2006p1742} and spin textures\cite{Vengalattore:2008p1759,Kronjager:2010p1758,Vinit:2013p1757}. However, the characteristic timescales in such systems are long, and equilibrium can be reached only under a very narrow range of conditions\cite{PhysRevA.84.063625}. Long-range order has also been introduced by using cavity photons\cite{Baumann:2010p1755} or lattice tilting\cite{Simon:2011p1574} to generate interactions, but also in a non-equilibrium context.

A distinct but complementary approach has focused on the design of lattices with complex dispersion relations.
The complexity of the band structure for cold atoms is now beginning to overcome the limitations of simple lattices in the ground band, and recent progress has seen more exotic lattices such as hexagonal\cite{SoltanPanahi:2011p1750} and kagome\cite{Jo:2012p1749}, as well as occupation of higher bands\cite{Wirth:2011p1490}.
One promising route toward more complicated band structures involves hybridizing the bands in a simple lattice by dynamically shaking the lattice\cite{Gemelke:2005p1741,Lignier:2007p1743,Struck:2011p1612,Struck:2012p1739}.
Experiments using this technique have created hybridized band structures with negative or near-zero tunneling coefficients\cite{Lignier:2007p1743}, or with multiple minima at high-symmetry points in the Brillouin zone\cite{Struck:2011p1612}.

\begin{figure}[htb]
\includegraphics[width=3.4 in]{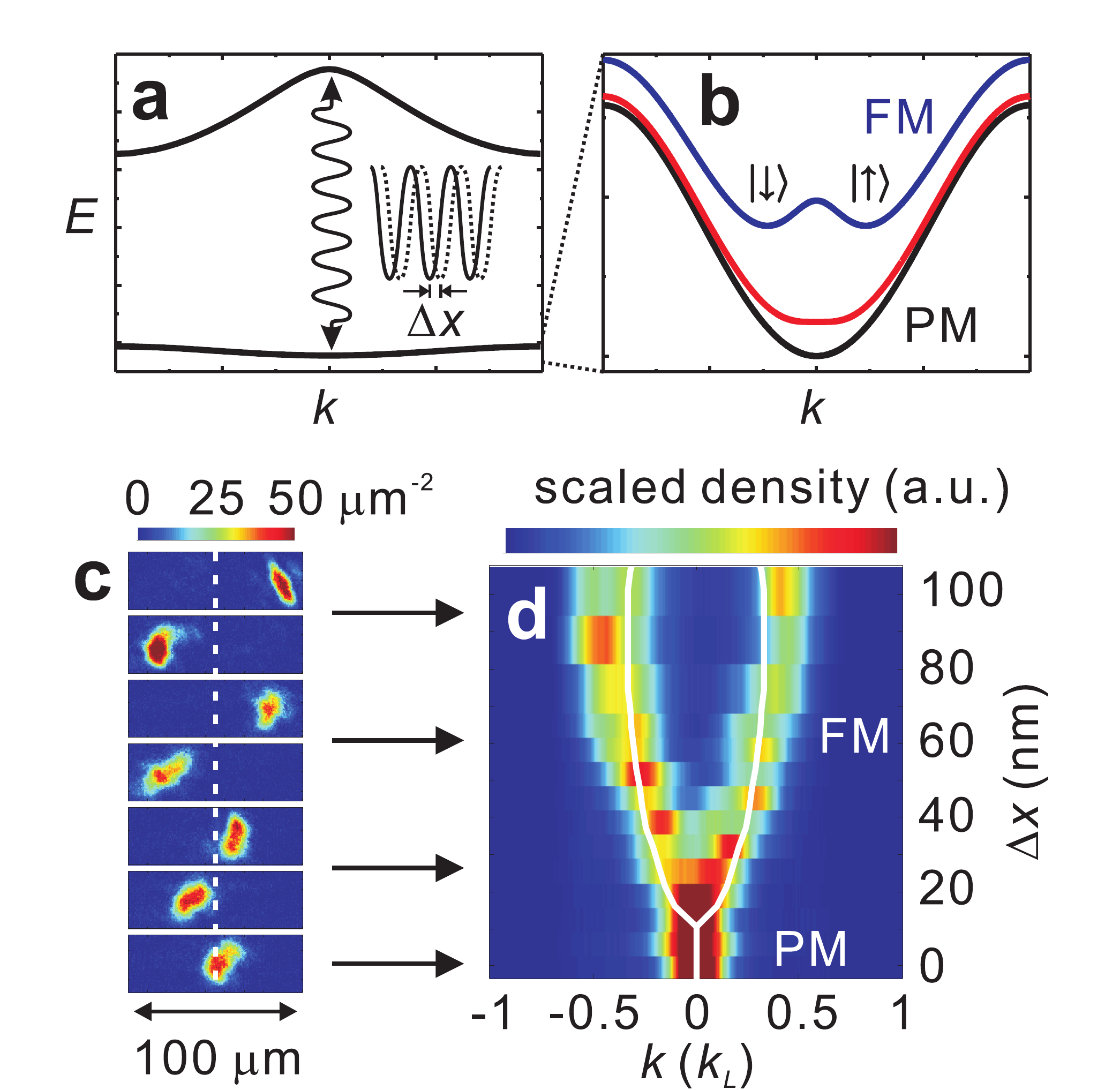}
\caption{\textbf{Ferromagnetic transition in a shaken optical lattice with double-well dispersion} (a) Dispersion $E(k)$ of the first two bands in an optical lattice, hybridized using near-resonant shaking. (b) Expanded view of the hybridized ground band in the paramagnetic case with no shaking (black), the ferromagnetic case with strong shaking (blue), and the critical case (red). (c) Single shot images (at $30$ ms TOF) of $\sim 25,000$ Cs atoms in the lattice with different shaking amplitudes. (d) Momentum distribution along the lattice direction as a function of peak-to-peak shaking amplitude $\Delta x$, averaged over 10 trials at each amplitude. The TOF position is used to determine the momentum in lattice units $k_L = 2\pi/\lambda_L$, where $\lambda_L = 1064\textrm{ nm}$ is the optical lattice wavelength. The white line is the calculated location of the dispersion minimum. We ramp on the shaking amplitude over 50 ms followed by an additional 50 ms of constant shaking.}\label{fig1}
\end{figure}

Our experiment is based on a cesium Bose-Einstein condensate (BEC) of ~25,000 atoms loaded into a one dimensional optical lattice (see Methods). Using lattice shaking near the ground-to-first-excited transition frequency we create a hybridized band structure with two distinct minima at momenta $k = \pm k^*$, with $k^*$ incommensurate to the lattice (see Fig. \ref{fig1}).
A similar dispersion has been obtained in the continuum by introducing Raman-dressed spin-orbit coupling\cite{Lin:2011p1764}, and there have been proposals for using this type of dispersion to generate spatially ordered phases\cite{Wang:2010p1773,Ho:2011p1772}.
Labeling the minima as spin-up and spin-down and treating the system as a two-mode BEC yields an effective Hamiltonian,
\begin{eqnarray}
\label{eqmain}
H & = & \sum_{\sigma = \uparrow, \downarrow} \epsilon_\sigma N_\sigma + \frac{g}{2}N^2_\uparrow + \frac{g}{2}N^2_\downarrow + 2gN_\uparrow N_\downarrow,
\end{eqnarray}
\noindent where $\epsilon_\sigma$ represents the single particle energy of each spin state, $N_\uparrow$ ($N_\downarrow$) is the number of up (down) spins, and $g=4\pi\hbar^2 a/mV$ is the interaction strength of the original gas in terms of the scattering length $a$, the mass $m$ and the effective trap volume $V$. The factor of two in the interspecies interaction arises from inclusion of both Hartree (direct) and Fock (exchange) interactions, and represents the large energy cost to support density waves when both modes are occupied.
Introducing the collective spin $\vec{J}$ representation\cite{Dalton:2011p1762}, we find the Hamiltonian for an easy-axis magnet,
\begin{eqnarray}
\label{eqspin}
H & = & \frac{\epsilon_\uparrow+\epsilon_\downarrow}{2}N + \frac{3g}{4} N^2 + bJ_z - g J_z^2
\end{eqnarray}
\noindent where $b = \epsilon_\uparrow - \epsilon_\downarrow$ is the effective field, $J_z = \frac{1}{2}\left(N_\uparrow - N_\downarrow\right)$ is the magnetization, and the first two terms are constants of motion.
For bosons with repulsive interaction $g > 0$, interactions between the two spins are strongly ferromagnetic.

To demonstrate this ferromagnetism, we ramp the amplitude of lattice shaking to tune the dispersion from one with a single minimum to one with two distinct minima. We perform absorption images after $30$ ms time-of-flight (TOF) with a magnetic gradient canceling the gravitational force of the earth; see Fig. \ref{fig1}c for sample images.
We also average over many shots to create a density histogram, shown in Fig. \ref{fig1}d.
When the lattice is removed abruptly, the atoms in different spin states are projected back to free particle states, giving us an effective Stern-Gerlach measurement. 
For no shaking up to a shaking amplitude of about $15\textrm{ nm}$ we observe a single, narrow momentum distribution centered at zero, consistent with a regular BEC.
As the shaking amplitude is increased further, we observe a bifurcation and the momentum distribution develops a two-peak structure. 
Note that this structure occurs after averaging over many shots to create the histogram; the majority of shots will feature all of the atoms in one state or the other, that is, fully magnetized samples.
Comparing to the calculated position of the minima from numerical diagonalization (white line in Fig. \ref{fig1}d, and see Supplementary Information), we find good overall agreement.
In the spin language, the transition from one to two minima corresponds to a paramagnetic (PM) to ferromagnetic (FM) transition.

\begin{figure}[htb]
\includegraphics[width=3.4 in]{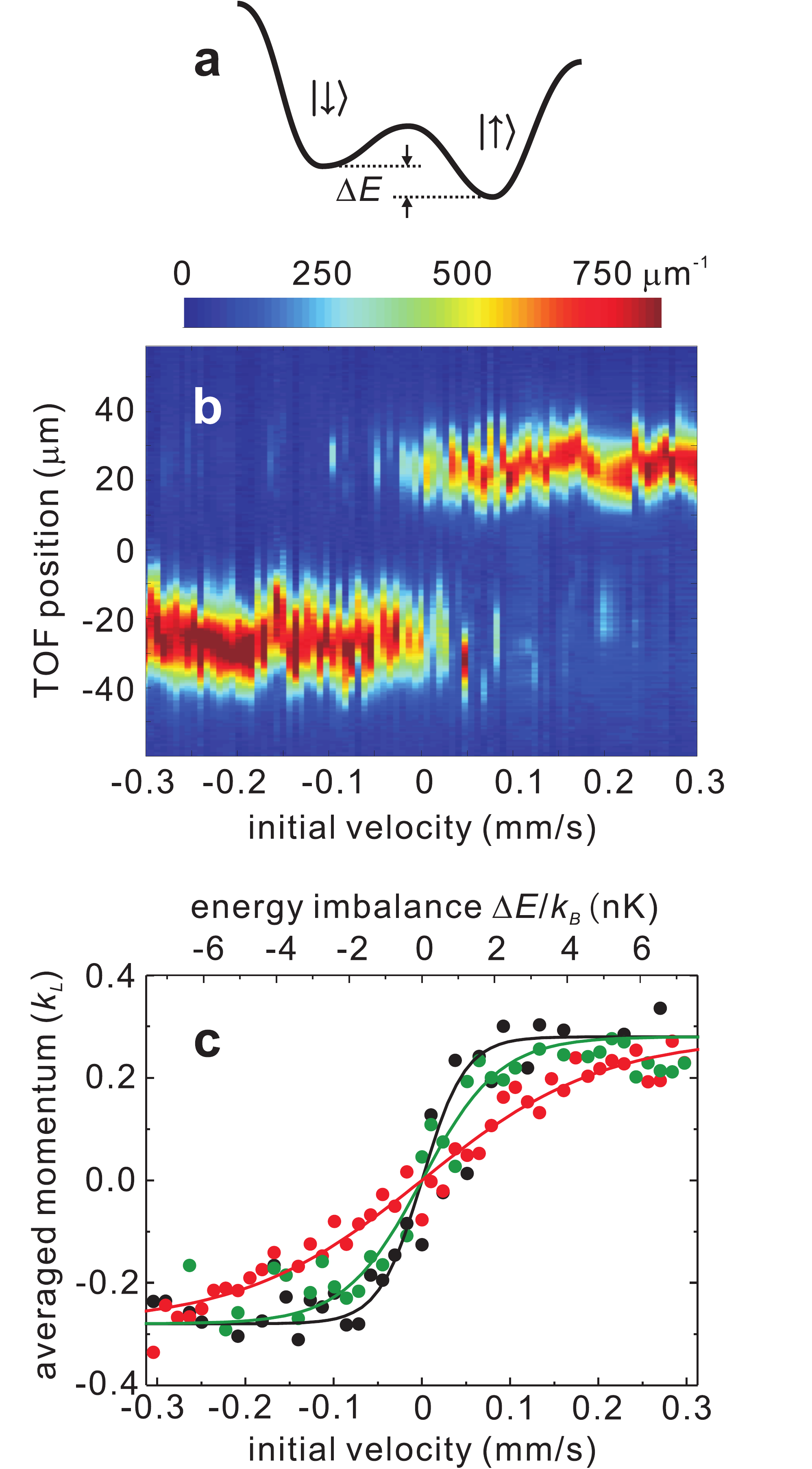}
\caption{\textbf{Sensitivity of the ferromagnetic transition to explicit symmetry breaking} (a) We control the energy imbalance $\Delta E$ with a small initial velocity $v$ of the atoms relative to the lattice. For short times the imbalance is given by $\Delta E = -b = 2\hbar k^*v$. (b) Average density profile along the lattice direction as a function of imbalance with $100\textrm{ ms}$ ramping time and scattering length $a = 35\textrm{ }a_0$. (c) Average momentum as a function of imbalance under three different conditions: ramping time $100\textrm{ ms}$ with $a = 35\textrm{ }a_0$ (black) or $a = 27\textrm{ }a_0$ (green), and ramping time $10\textrm{ ms}$ with $a = 35\textrm{ }a_0$ (red). The solid lines are fits to a thermal distribution, with effective temperatures $T_{\rm eff}$ of $0.7$, $1.2$, and $2.9\textrm{ nK}$. The sample has temperature $T = 7\textrm{ nK}$ and chemical potential $\mu \sim 20\textrm{ nK}$. Atoms were held at constant shaking amplitude $\Delta x = 65 \textrm{ nm}$ for $100\textrm{ ms}$ following the ramp.\label{fig2}}
\end{figure}

The complete magnetization of the sample above a critical shaking amplitude demonstrates a spontaneous symmetry breaking process.
We investigate this process by testing its sensitivity to an explicit symmetry breaking term $b J_z$, see Fig. \ref{fig2}.
This is realized by providing the condensate a small initial velocity $v$ relative to the lattice that acts as a synthetic field $b = -2\hbar k^* v$, where $2\pi\hbar$ is the Planck constant (see Supplemental Information).
For spontaneous symmetry breaking we expect to fully magnetize the sample even for symmetry breaking $b$ much less than our temperature scale or chemical potential.
To quantify the sensitivity we assume atoms populate the two spin states according to a Boltzmann distribution with an effective temperature $T_\textrm{eff}$.
When the lattice shaking, and thus ferromagnetism, is ramped on slowly over $100\textrm{ ms}$ we find an effective temperature of $0.7\textrm{ nK}$, well below the actual temperature of $7\textrm{ nK}$ and chemical potential $\sim 20\textrm{ nK}$.
This extreme sub-thermal sensitivity shows that our system is driven into a fully ferromagnetic state by spontaneous symmetry breaking. When ferromagnetism is ramped on more quickly, the sensitivity is reduced, which is the expected behavior of a quenched ferromagnet.
We can identify the importance of interactions to the symmetry breaking by changing the bias magnetic field to vary the scattering length via Feshbach resonance\cite{RevModPhys.82.1225}.
When the scattering length is reduced from $35\textrm{ }a_0$ to $27\textrm{ }a_0$, with $a_0$ the Bohr radius, we observe a less sensitive transition, which confirms that spin interactions depend on the scattering length.

\begin{figure}[htb]
\includegraphics[width=3.4 in]{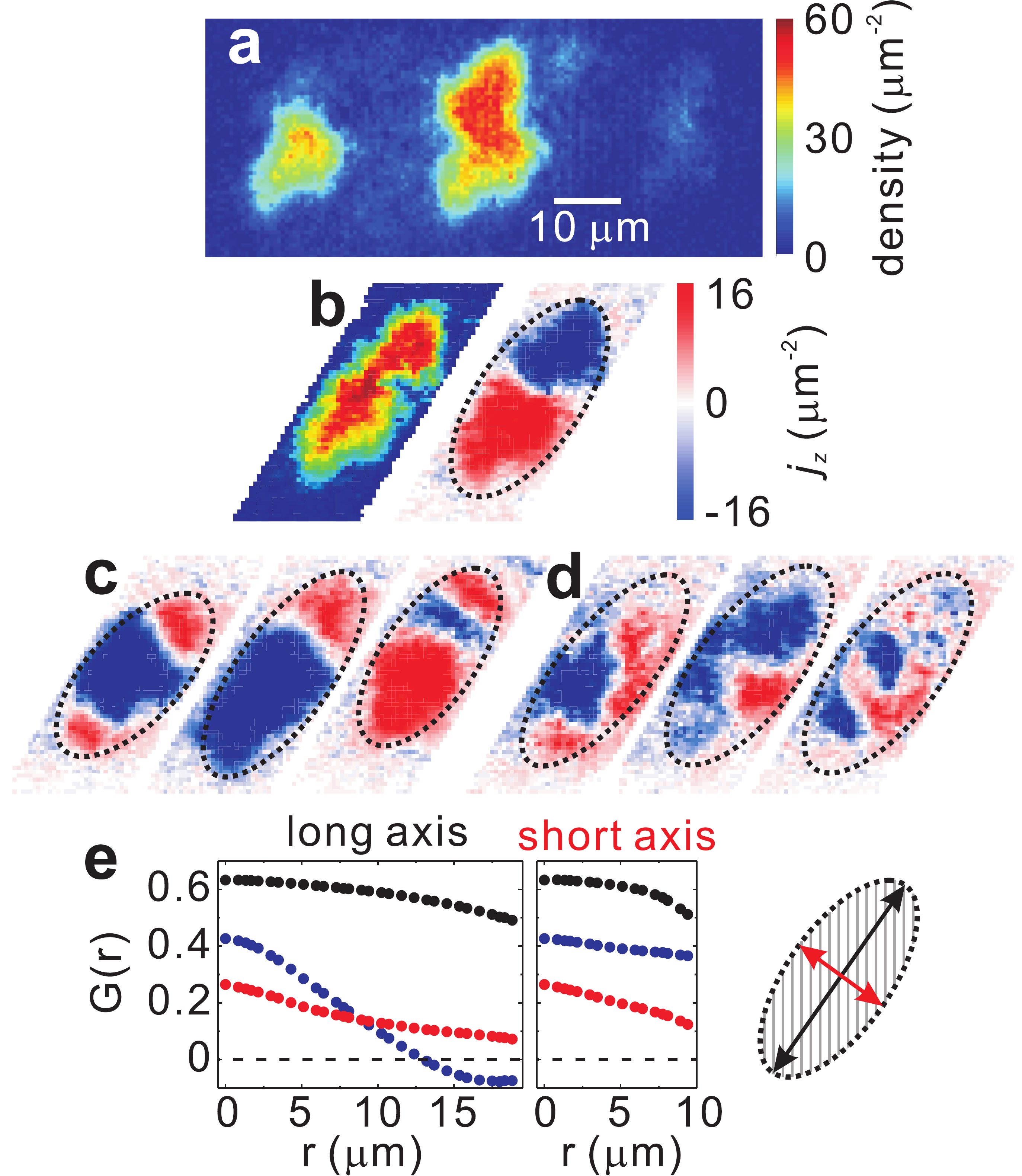}
\caption{\textbf{Ferromagnetic domains and spin correlations} (a) Image taken after 5 ms TOF, showing lattice Bragg peaks. (b) Reconstructed density and magnetization using Bragg peaks (see Supplementary Information). (c) Three representative magnetization images with $100\textrm{ ms}$ ramping time. (d) Three representative magnetization images with $10\textrm{ ms}$ ramping time. (e) Density-weighted magnetization correlator $G(r)$ along the long and short trapping directions: The average of $\sim 10$ single domain samples with 100 ms ramping time (black), $\sim 10$ multi-domain samples with $100\textrm{ ms}$ ramping time (blue), and $\sim 20$ multi-domain samples with $10\textrm{ ms}$ ramping time (red). Atoms were held at constant shaking  amplitude $\Delta x = 65\textrm{ nm}$ for $100\textrm{ ms}$ following the ramp.
}\label{fig3}
\end{figure}

When a ferromagnet is cooled in the absence of an external bias field, domain formation is expected.
Here too we observe that with the smallest symmetry breaking applied, or rapid ramping of the ferromagnetic interaction, we can form domains. Once the shaking has ramped on, and the domains have formed, the confining potential or other sources can no longer move atoms across the barrier to the other spin state.
Therefore, total magnetization (i.e. total quasimomentum) will be conserved.
Figure \ref{fig3}a shows the domain structure at 5 ms TOF for a typical situation when ferromagnetism is ramped on slowly (over $100$ ms).
A more detailed reconstruction of the original domain structure can be accomplished by taking advantage of the information in the Bragg peaks (see Supplementary Information).
Such domains are further proof of the symmetry breaking nature of our system.
As one might expect for a ferromagnet, the nature of these domains depends on the conditions in which they were formed.
When ferromagnetism is ramped on slowly over $100$ ms, we observe larger domains, with boundaries typically oriented in the same direction (Fig. \ref{fig3}c).
When the ramping is done as a quench, over $10$ ms, we observe a greater number of smaller domains, with less predictable orientation (Fig. \ref{fig3}d). Our result is consistent with the Kibble-Zurek mechanism in the sense that faster ramps yield shorter range correlations.

To quantify difference in domain size and shape, we compute the density-weighted magnetization correlator\cite{Sadler:2006p1742},
\begin{eqnarray}
G(\delta \mathbf{r}) = \frac{\left\langle\int j_z(\mathbf{r})j_z(\mathbf{r}+\delta \mathbf{r})d\mathbf{r}\right\rangle}{\left\langle\int n(\mathbf{r})n(\mathbf{r}+\delta \mathbf{r})d\mathbf{r}\right\rangle},
\end{eqnarray}
where $n$ and $j_z$ denote number and magnetization densities, and angle brackets denote an average over multiple trials. We distinguish between single and multiple domain samples (see Supplemental Information). For fully polarized domains we expect $G(0) = 1$, however, we obtain $G(0) = 0.6$ for single-domain samples, which can be explained by a $10\%$ uncertainty in state identification. $G(0)$ is even lower for samples with domains due to the observed domain wall size, which is limited by atom dynamics during the 5 ms TOF. Along the short trap axis, the correlations in samples with slow ramping are both stronger (indicating fewer domain boundaries) and longer range (indicating larger domains) compared with quenched samples. In quenched samples the correlations are roughly isotropic due to the random orientation of domains. In samples with slow ramping and multiple domains, the correlation along the long trap axis drops off abruptly at about $10 \textrm{ }\mu\textrm{m}$ or 20 lattice spacings, see Fig. \ref{fig3}. Our analysis of $G(r)$ demonstrates that long-range spin correlations can be established, and that domains boundaries prefer to align along the short trap axis when ferromagnetism is turned on slowly.

\begin{figure}[htb]
\includegraphics[width=3.4 in]{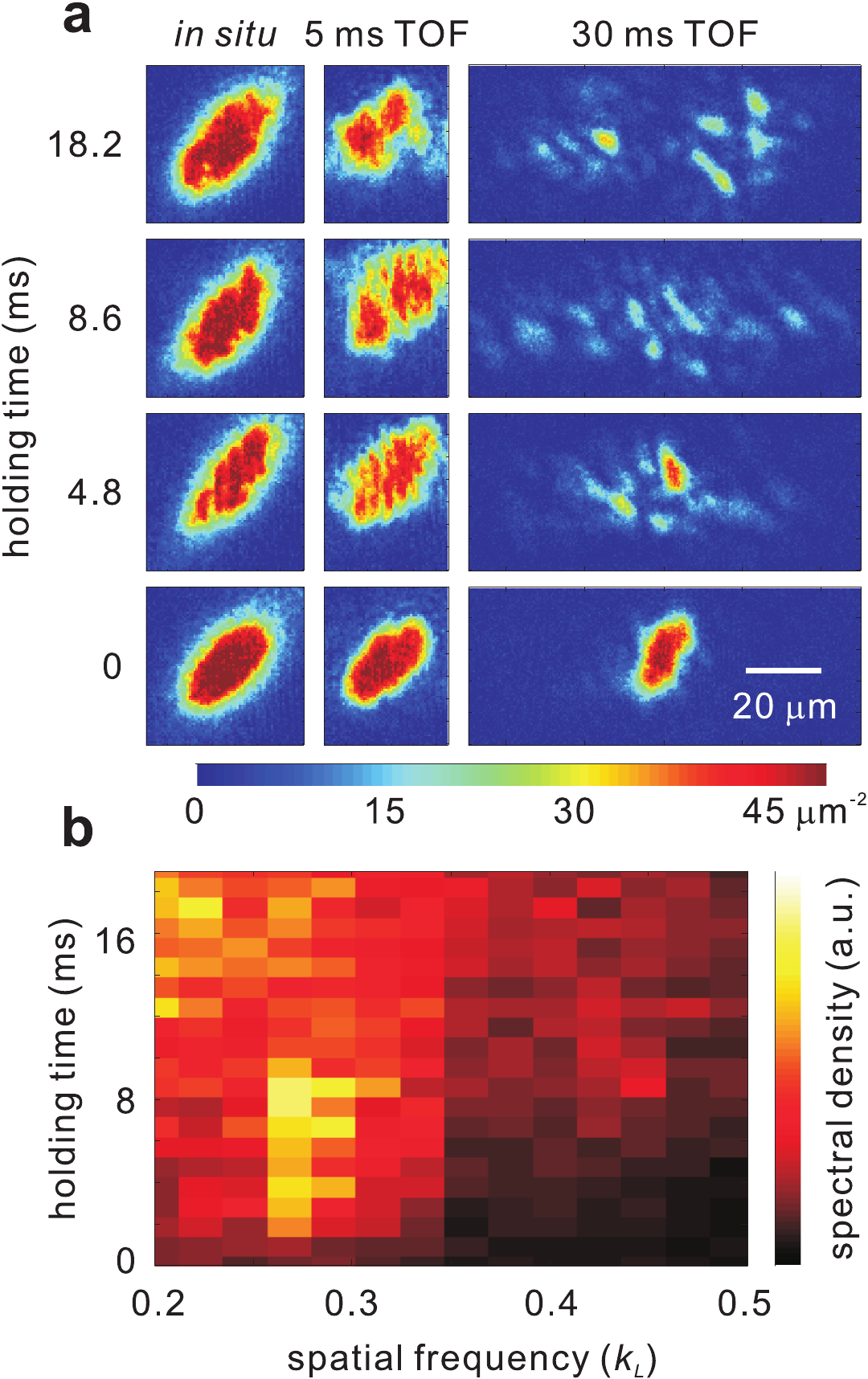}
\caption{\textbf{Quench dynamics and magnetic domain interference} (a) Single shot images taken \textit{in situ} and with 5 ms and 30 ms TOF, at several hold times following a 5 ms quench into the ferromagnetic state. (b) Spatial power spectrum along the lattice direction from images with 5 ms TOF, averaged over 20 shots. A peak appears at $k^* = 0.27 k_L$ for the first 10 ms. The shaking amplitude was $\Delta x = 65 \textrm{ nm}$.
}\label{fig4}
\end{figure}

To fully investigate the emergence of domains from a single mode condensate, we measure the spatial and momentum distribution of the atoms after a sudden ($5\textrm{ ms}$) quench across the ferromagnetic transition. Figure \ref{fig4}a shows images at various hold times following the quench and for different TOF, revealing that immediately following the quench the atoms have not yet moved appreciably from their original momentum distribution, and therefore in unstable equilibrium at zero momentum.
Over the course of about $10$ ms, the atoms displace from this maximum into the minima on either side in a complex and dissipative manner, eventually completely depopulating the zero momentum state (see Fig. \ref{fig4}a). The dissipative dynamics indicate that energy must flow into other degrees of freedom, for example the kinetic energy in the transverse (non-lattice) directions.
Observation of fast mixing between the spin and motional degrees of freedom demonstrates that our spin-spin interactions are strong and will drive the system towards equilibrium on short timescales.

Given the quantum nature of our magnetic domains, which are characterized by complex order parameters $e^{ik^*x}\Psi(x)$ and $e^{-ik^*x}\Psi(x)$, where $\Psi(x)$ is the bosonic field operator, we expect spatial interference if they were made to overlap.
We do indeed see interference at intermediate times hold times for \textit{in situ} and $5\textrm{ ms}$ TOF images. Figure \ref{fig4}b shows the fast Fourier transform of the atomic density averaged over multiple 5 ms TOF images, showing a peak at wavevector $0.27 k_L = 0.9k^*$, where $k_L = 2\pi/\lambda_L = 2\pi/{1064\textrm{ nm}}$ is the lattice momentum.
This signal, at half the expected wavevector for interference between the two domains, is consistent instead with interference between either domain and the remnant population at zero momentum.
The interference grows in strength as the hold time increases and the system relaxes from the quench, reaching a peak at $\sim 10$ ms.
This supports our interpretation, as at longer times the system nears equilibrium, domains have formed and there is no remnant population at zero momentum.
Because our three dimensional condensate is thicker than the depth of focus of the imaging system, we lack the resolution to detect interference at $2k^*$.
We also note that the interference is weaker for \textit{in situ} images compared with those taken at 5 ms TOF.
This suggests that as the condensate begins to relax toward the two minima, it has already begun to break up in real space to reduce density corrugation.
With a time-of-flight image, the domains pass over one and other, allowing us to visualize the quantum inference more clearly.

In conclusion, we have demonstrated a novel method for creating and observing long-range magnetic order using a double-well band structure created by near-resonant lattice shaking. We are able to modify the dispersion quite significantly with only minimal heating. With increasing shaking strength we can realize a paramagnetic to ferromagnetic transition. The ferromagnetic state can support multiple domains, and is extremely sensitive to the symmetry breaking field. Using lattice shaking to tune band structure has important implications for the simulation of various ordered states in solid systems, where Fermi surface shape and topology can play a very important role. The same double well used here would have a nested Fermi surface and be expected to undergo a charge density wave transition\cite{Shankar:1994p1745}, for example. Furthermore, the near-resonant shaking technique is easily extendable to two or three dimensions, or other atomic species, which can be Fermionic and/or contain multiple accessible internal states. Thus near resonant shaking opens the door to a variety of exciting possibilities for quantum simulation.

\section{Methods}
\subsection{Lattice loading}
Our experiment begins by evaporating and loading a $^{133}\textrm{Cs}$ BEC into a three dimensional optical dipole trap with trapping frequencies of $8.6$, $19.1$, and $66.9$ Hz in three directions, with the tightest trapping in the direction of gravity and imaging\cite{Hung:2008p1751}.
The atoms are then loaded into a one-dimensional optical lattice at $35^\circ$ to the in-plane trapping directions, where the final atom number is between 20,000 and 30,000 at a temperature of 7 nK.
After the atoms are loaded into the optical lattice, a sinusoidal shaking is turned on with a linear ramp of between $5$ and $100$ ms. After the shaking is ramped on, we shake the atoms for 50-100 ms before performing an \textit{in situ} image or extinguishing all lattice and trapping light for a time-of-flight image. Moderate heating is observed during the lattice shaking, resulting in a lifetime of 1 s.
Our optical lattice is formed by reflecting one of the dipole trap beams back on itself after passing through two oppositely oriented acousto-optic modulators (AOMs).
The lattice modulation is accomplished by frequency modulating the driving radio frequency (around a carrier of $80$ MHz) for the paired AOMs, which changes the relative phase, and therefore the optical path length, between the AOMs (see Supplemental Information).

\subsection{Lattice shaking}
To realize a double-well dispersion, we use a periodic shaking of the optical lattice at a frequency near the ground band to first excited band transition at zero quasi-momentum.
This shaking allows the two bands to mix, creating a competition between the positive curvature of the ground band and the negative curvature of the excited band, as shown in Fig. \ref{fig1}.
For our experiments we use a laser wavelength $\lambda_L = 1064\textrm{ nm}$ (lattice spacing $532\textrm{ nm}$) and lattice depth $V = 7.0 \textrm{ }E_R$, where $E_R = h^2/2m\lambda_L^2$ is the lattice recoil energy.
This yields a zero momentum band gap of $5.0\textrm{ }E_R$.
We apply the shaking at a slightly blue-detuned frequency of $7.3 \textrm{ kHz} = 5.5\textrm{ }E_R/h$, which gives the least heating when the double-weel dispersion is formed.
The solid black curves in Fig. \ref{fig1}b show the lowest two bands without shaking in the dressed atom picture.
To confirm that the bands will mix to create a double-well potential, we have numerically computed the hybridized Floquet states for several different shaking amplitudes and the results agree well with the experiment (see Supplemental Information).

\section{Acknowledgements}
We thank Chung-Kuan Lin for assistance in the early stages of the experiment. We acknowledge useful discussions with N. Gemelke, I. Spielman, A. Ran\c{c}on, H. Zhai, and G. Baym. This work was supported by NSF MRSEC (DMR-0820054), NSF Grant No. PHY-0747907 and ARO Grant No. W911NF0710576 with funds from the DARPA OLE Program.

\section{Author contributions}
L.-C.H. performed the experiments. L.-C.H. and C.V.P. analyzed the data and C.V.P. wrote the manuscript. C.C. supervised. We declare no competing financial interest.

\bibliographystyle{naturemag}
\bibliography{cvppapers}

\end{document}


\title{Supplement to ``\textit{In situ} observation of strongly interacting ferromagnetic domains in a shaken optical lattice''}
\author{Colin V. Parker}
\affiliation{The James Franck Institute and Department of Physics, University of Chicago, Chicago, IL 60637, USA}
\author{Li-Chung Ha}
\affiliation{The James Franck Institute and Department of Physics, University of Chicago, Chicago, IL 60637, USA}
\author{Cheng Chin}
\affiliation{The James Franck Institute and Department of Physics, University of Chicago, Chicago, IL 60637, USA}
\affiliation{Enrico Fermi Institute, University of Chicago, Chicago, IL 60637, USA}
\date{\today}



\maketitle

\section{Numerical calculations of the hybridized band}
The single-particle physics of an atom in a (1D) time-dependent optical lattice formed by retroreflected light of wavelength $\lambda_L$ is governed by the Hamiltonian
\begin{eqnarray}
\label{eqn1}
H(t) & = & E_R\left[\frac{\lambda_L^2}{4\pi^2}\frac{\partial^2}{\partial x^2} + V\cos\frac{4\pi}{\lambda_L} \left(x-x_0(t)\right)\right],
\end{eqnarray}
where $E_R = h^2/2m\lambda_L^2$ is the recoil energy, $\lambda_L/2$ is the lattice constant, $V$ is the lattice depth in terms of recoils, and $x_0(t)$ is the time-dependent lattice offset.
For periodic lattice offsets $x_0(t+T) = x_0(t)$, the Hamiltonian is characterized by temporally and spatially periodic Floquet states.
In analogy with spatially periodic Bloch states, which define momentum only up to an overall lattice momentum, the Floquet states define energy only up to an overall energy $E = h/T$, corresponding to the absorption or emission of one quantum of energy at the shaking frequency.
We break the Hamiltonian (\ref{eqn1}) into discrete time steps and numerically diagonalize in order to determine the minima of the hybridized bands.

\subsection{Initial velocity and effective field}
To induce an effective energy imbalance we use controlled initial velocity changes to correct for random variations in laboratory conditions.
Over times much shorter than the trapping frequency, the effective energy imbalance of a given initial velocity can be quantified by performing a Galilean transformation to the atomic reference frame.
This modifies the dispersion relation $E(k) \rightarrow E(k)-\hbar k v$, where $v$ is the initial velocity.
For small $v$ this will split the spin states by $\pm\hbar k^* v$, where $\pm k^*$ corresponds to the positions of the minima.
Over longer periods this approximation breaks down, but it remains true that initial velocity breaks the symmetry between the two states and transformation provides an approximate energy scale.
Fortunately, laboratory conditions change on long enough timescales that several successive shots can be acquired under similar conditions.
We can alter the initial velocity around the random value by changing the relative beam powers of the trapping beams, which moves the trap center slightly and alters the initial velocity.
The initial velocity changes are quantified by observing the change in momentum of a cloud with no lattice shaking during a series of control experiments taken between successive data shots.
Because our random variations are larger than our sensitivity, we have performed several scans under each set of conditions ($\sim 40$) and adjusted the $50\%$ point of each scan to zero.
If the midpoint of the transition were characterized by random statistics, with a $50\%$ chance of complete-up and $50\%$ chance of complete down, this would tend to bias us toward a sharper transition by a small amount ($\sim 1$ nK).
However, as the midpoint of the transition involves domain formation, this bias will be a much smaller effect.
To determine the effective temperature, we assume a Boltzmann distribution, where the average momentum $M$ in terms of initial velocity $v$ would be
\begin{eqnarray}
\label{eqn2}
M(v) = \hbar k^*\tanh\left(\frac{\hbar k^* v}{k_B T_\textrm{eff}}\right).
\end{eqnarray}

\begin{figure}[ht!]
\includegraphics[width=2.9 in]{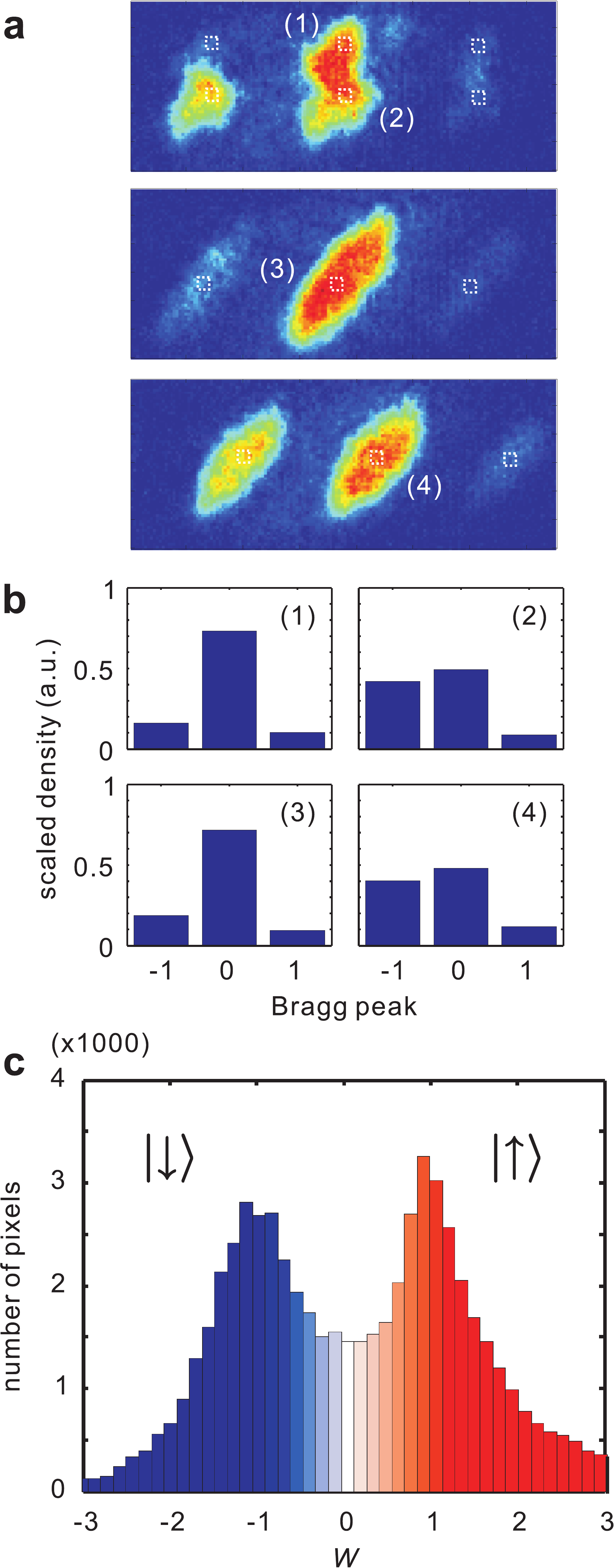}
\caption{\textbf{Domain reconstruction} (a) Images at 5 ms TOF for a multi-domain sample, and two single domain samples. (b) Vectors $\vec{w}$ signifying relative weights between the 0, 1, and -1 order Bragg peaks, corresponding to the regions in (a) identified by white boxes. (c)Histogram (over many images) of pixel-wise projection onto $\vec{w}_\uparrow-\vec{w}_\downarrow$. The two peaks occur because the large majority of the samples are pure spin-up or spin-down. \label{figs:domain}}
\end{figure}

\section{Domain reconstruction}
In images taken with $5$ ms TOF the domain structure is visible in the center of the image, as portions of the cloud move in opposite directions.
On either side of the main cloud are higher momentum Bragg peaks induced by the lattice, which has been abruptly turned off along during the expansion.
True \textit{in situ} images cannot distinguish the domains, while for longer TOF the shape of the domains is distorted too much during the expansion.
To leading order atoms in both spin states respond to the lattice shaking with the same phase, so the physical motion (in the lab frame) for either must be the same.
If the atoms are released when traveling to the right, the spin-up atoms can have mostly momentum $k^*$, with only small amounts of higher Bragg peaks at $k^* \pm 4\pi/\lambda_L$.
By contrast, spin-down atoms must have a large fraction at $-k^* + 4\pi/\lambda_L$ in order to be physically moving right.
Thus, the relative strength of the center and side two Bragg peaks forms a signature that identifies the spin state of each section of the image.
We define a three-dimensional vector which is proportional to the density at each Bragg peak, $\vec{w} = (n_L, n_0, n_R)$ (Fig. S\ref{figs:domain}). The vector is defined pixel-wise and is normalized to sum to one. Using images which can be identified clearly as fully spin-up or spin-down, we determine typical vectors $\vec{w}_\uparrow$ and $\vec{w}_\downarrow$. Then when analyzing multi-domain images, we compute the population fraction of different components in each pixel by projecting the vector along the $\vec{w}_\uparrow-\vec{w}_\downarrow$ axis. A histogram of the resulting projection, which we denote as $W$, is shown in Fig. S\ref{figs:domain}c. We associate the peaks of this histogram with spin-up and spin-down, and values in between with fractional population according to the value of $W$. Values of $W$ beyond the histogram peaks at assumed to be fully magnetized. The vectors $\vec{w}$ as well as the histogram peaks depend on the phase of the lattice modulation at the TOF release time, and we calibrate both under different experimental conditions. After determining the pixel-wise spin density of a $5\textrm{ ms}$ TOF image, we shift each image back by the expected travel distance during the TOF (left or right, depending on spin) and reconstruct the number and magnetization densities at the time of release.

\section{Correlation analysis}
In the correlation analysis, we distinguish multi-domain samples as those with a nearly equal amount of population in different states. We choose those where the sum of magnetization divided by the sum of density is smaller than 0.275, where $0.5$ is the theoretical value for a fully polarized sample. This corresponds roughly to having less than $77.5\%$ of the sample in the same state. For the analysis of fully magnetized sample, we select images with this ratio larger than 0.305, or more than $80.5\%$ in the same state. Although this threshold may seem low, our determination of the magnetization has associated error, and empirically these numbers correctly distinguish the single- and multi-domain cases.

\section{Phase-modulating the lattice}
\begin{figure}[ht]
\includegraphics[width=3.4 in]{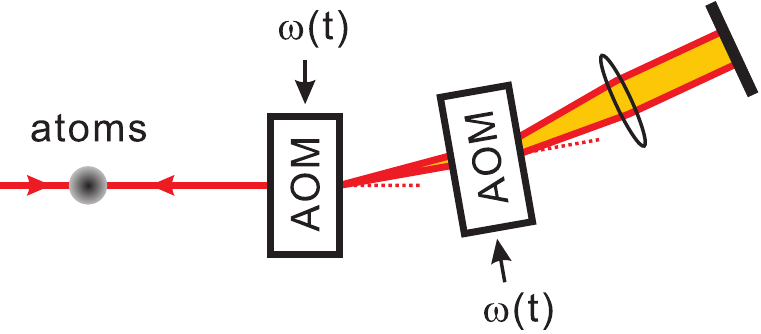}
\caption{\textbf{Lattice phase modulation setup} The lattice beam passes through two AOMs operating at the same frequency (from the same source), passing each AOM twice. The total frequency shift is zero so the return beam can make a lattice with the incoming beam. The lens is positioned so that the AOM frequency can change while maintaining the retroreflection condition. When the AOM frequency changes,  a phase shift develops in the beam path between the two AOMs leading to a change of the optical lattice phase.\label{figs:opt_setup}}
\end{figure}

Our optical setup is shown in Fig. S\ref{figs:opt_setup}. The lattice shaking is accomplished by sinusoidally varying the phase of the retro-reflected beam. The beam passes through two acousto-optic modulators (AOMs) which, combined, leave the phase unaffected, whereupon it is focused by a mirror, then reflected to pass through the same AOMs again, for a total of four AOM passes\cite{xibo_thesis}. When the AOM frequency is changed, the relative phase of the acoustic waves in the two AOMs changes, leading to variable total phase change on reflection. The sinusoidal modulation of the phase is then done by frequency modulation of the AOM signal.
We calibrate the phase modulation by measuring the momentum kick acquired by a step function lattice displacement. A null result occurs for steps with are multiples of $2\pi$, allowing us to calibrate in terms of the lattice spacing.

\bibliographystyle{naturemag}
\bibliography{cvp,cvppapers}